\documentclass[aps,prl,twocolumn,groupedaddress]{revtex4}
\usepackage{graphicx}
\begin{document}

\title{Two-Sided Gravitational Mirror: Sealing off Curvature Singularities}

\author{Aharon Davidson and Ben Yellin}
\email[Email: ]{davidson@bgu.ac.il}

\affiliation{Physics Department, Ben-Gurion University,
Beer-Sheva 84105, Israel}

\date{January 1, 2012}

\begin{abstract}
A gravitational mirror is a non-singular finite redshift surface which
bounces all incident null geodesics.
While a white mirror (outward bouncing) resembles 't Hooft's brick
wall, a black mirror (inward bouncing) offers a novel mechanism for
sealing off curvature singularities.
The geometry underlying a two-sided mirror is characterized by
a single signature change, to be contrasted with the signature flip
which governs the black hole geometry. 
To demonstrate the phenomenon analytically, we first derive an exact,
static, radially symmetric, two-sided mirror solution, which asymptotes
the massless BTZ black hole background, and then probe the local
structure of a massive mirror.
\end{abstract}

\pacs{04.20.Dw, 04.60.Kz, 04.20.Jb}

\maketitle

A signature change in the metric of spacetime plays a fundamental role
in general relativity and beyond.
First and foremost in black hole physics, where crossing the horizon is
accompanied by the signature flip (= double signature change)
$\{-,+,...\}_{out}\leftrightarrow \{+,-,...\}_{in}$, with the dots representing
unchanged spacelike signatures.
Still, as is evident from the finiteness of the various curvature scalars,
the horizon itself is a non-singular surface.
The cosmological example is provided by Hartle-Hawking 'no-boundary'
proposal \cite{NoBoundary}, originally invoked in an attempt to bypass
the classical Big Bang singularity.
The quantum creation of the universe is then described in terms of
a smooth Euclidean to Lorentzian transition governed by the single
signature change $\{+,+,...\}_E\leftrightarrow \{-,+,...\}_L$.
It has even been argued that, for discontinuous signature-changing
metrics \cite{SignatureChange}, one can derive the field equations from
a suitable variational principle. 
This paper is devoted to a signature change of a third type, namely
$\{-,+,...\}_{out}\leftrightarrow \{-,-,...\}_{in}$, introducing a novel black
configuration, to be referred to as a two-sided mirror, a non-singular
finite redshift surface which bounces all incident null geodesics.

Let our starting point be the action principle
\begin{equation}
	{\cal I}=-\int \left( \phi {\cal R}+V(\phi) \right)\sqrt{-g}~{d^3}x ~.
\end{equation}
Working in 3-dim is just a matter of choice, allowing us to
make our point by means of an exact analytic solution; the 4-dim
conclusions although very similar, are based at this stage, on numerical
plotting and analytic probing at the near mirror zone.
Similar actions have been successfully used in 2-dim black
hole physics \cite{2dim}.
The primary role of the scalar potential $V(\phi)$ is to ensure a stable
general relativistic vacuum solution.
This can be done by letting the dilation field $\phi(x)$ acquire a constant
vacuum expectation value $\langle\phi\rangle=v=G^{-1}$, with $G$
denoting the 3-dim Newton constant.
Supplementing the action by Brans-Dicke kinetic $\omega$-term is
optional, but even in its absence (without losing generality), the dilation
field is still dynamical.
Associated with the above action are the field equations
\begin{eqnarray}
	&\phi G_{\mu\nu}
	+\left(\phi _{;\mu\nu}-g_{\mu\nu}g^{\alpha \beta }
	\phi _{;\alpha \beta }\right)=\frac{1}{2}g_{\mu\nu} V(\phi)  ~, & \\ 
	& \displaystyle{{\cal R}+\frac{dV(\phi)}{d\phi}=0 ~.} & 
\end{eqnarray}
By tracing the former gravitational field equations, and then substituting
the Ricci scalar into the latter equation, one infers that the Klein Gordon
equation of the dilation field, namely
$g^{\mu\nu}\phi_{;\mu\nu}=V^{\prime}_{eff}(\phi)$,
is governed by the derivative of the effective potential
\begin{equation}
	V_{eff}(\phi)=\frac{1}{4}\int\left(\phi V^{\prime}(\phi)
	-3V(\phi)\right)d\phi ~.
\end{equation}
The simplest choice by far, namely a linear $V(\phi)$,
(i) Does not have an equivalent $f({\cal R})$ gravity \cite {f(R)}
representation,
(ii) Dictates a constant Ricci curvature solution, and
(iii) Is sufficient for inducing a quadratic
$V_{eff}(\phi)$, that is
\begin{equation}
	V(\phi)=\frac{6}{\ell ^2}\left(\frac{2v}{3}-\phi \right) 
	~\Longleftrightarrow~
	V_{eff}(\phi)=\frac{3}{2\ell^2}(\phi-v)^2 ~.
\end{equation}
Stability, meaning an effective potential bounded from below, then
requires $\ell^2>0$, which signals a necessarily negative cosmological
constant $\Lambda=-\ell^{-2}$.
Note that adding a conformal piece  $\delta \phi^3$ to $V(\phi)$
keeps the effective potential $V_{eff}(\phi)$ unchanged, with
the sole effect being then a shift $\frac{1}{2}\delta v^2$ in the size of
the cosmological constant.

The most general line element admitting
$\partial/\partial t,\partial/\partial \varphi$ Killing vectors takes the
generic form
\begin{equation}
	ds^2=-e^{\nu(r)}dt^2+e^{\lambda(r)}dr^2
	+r^2\left(d\varphi+A(r)dt \right)^2 ~.
\end{equation}
For the sake of the present paper, however, the radially symmetric
 case $A(r)=0$ will do.
Re-arranging the field equations, we face a set of three differential
equations, two of which are second order, and one of which is
independent of the scalar potential, namely
\begin{eqnarray}
	&&\phi ''-\frac{1}{2}\left(\lambda'+\nu'\right)
	\left( \phi '+\frac{\phi }{r}\right)=0 ~,\\
	&&\phi '+\frac{\left(\nu '-\lambda '\right)}{2}\phi
	-\frac{3r}{\ell ^2} e^{\lambda }
	\left(\phi-\frac{v}{3}\right)=0 ~,\\
	&&\phi ''+\left(\frac{1}{r}+\frac{\nu '-\lambda '}{2}\right)\phi '
	-\frac{3}{\ell ^2}e^{\lambda }(\phi -v)=0 ~.
\end{eqnarray}
Constant $\phi(r)=v$ is obviously a solution, the vacuum solution
in fact.
The associated general relativistic metric
\begin{equation}
	ds^2_{BTZ}=-B(r)dt^2
	+\frac{dr^2}{B(r)}
	+r^2 d\varphi^2 ~,
	\label{BTZ}
\end{equation}
where $B(r)=-M+\ell^{-2}r^2$, constitutes the BTZ black hole metric
\cite{BTZ}.
Apart from its negative cosmological constant $\Lambda=-\ell^{-2}$,
the BTZ solution is characterized by a positive ADM mass $M\geq 0$.
Intriguingly, the AdS limit is achieved for $M=-1$,
indicating a mass gap of $\Delta M=1$ from the black hole continuum
states.
The outer and the inner horizons, as well as the infinite redshift surface,
which characterize the more general $J\neq 0$ metric, merge in this
case, at $h=\sqrt{M\ell^{2}} $, into a single event horizon.
For the sake of the present paper, it is worth emphasizing the well known
fact that although crossing the horizon is associated with a signature
flip, namely $(-,+,+)_{out}\leftrightarrow(+,-,+)_{in}$,
the horizon surface is nonetheless free of any curvature singularity.

In this paper, however, we are interested in the  non general relativistic $\phi(r)\neq v$ solution.
Following the Klein Gordon equation, a tiny deviation from the vacuum
expectation value, that is
$\phi(r)\simeq v\left(1+\delta(r)\right)$, is given by
\begin{equation}
	\delta(r)=\frac{s^3}{r^{3}}+\frac{r}{r_{0}}~,
	\label{asymptotic}
\end{equation}
where $s^3,r_0^{-1}$ are two constants of integration.
The choice $r_{0}\rightarrow \infty$ is then dictated by the boundary
requirement of asymptotically approaching the general relativistic $M=0$
BTZ background.
The corresponding full large-$r$ expansion, subject to the BTZ boundary condition
$\phi(r)\rightarrow v$ at infinity, is explicitly given by
\begin{eqnarray}
	&& e^{\nu(r)}=-M+\frac{r^2}{\ell^2}+\frac{2s^3M}{r^3}
	+\frac{3s^3M^2\ell^2}{2r^5}
	+...  ~, \nonumber \\
	&& e^{-\lambda (r)}=-M+\frac{r^2}{\ell^2}+\frac{8s^3}{r\ell^2}
	+\frac{3s^3M}{r^3}+...  ~,\\
	&& \phi(r)=v\left(1+\frac{s^3}{r^3}+\frac{3s^3M\ell^2}{4r^5}
	-\frac{20s^6}{7r^6}+... \right) ~. \nonumber
	\label{expansion}
\end{eqnarray}
The coefficients of the various $\displaystyle{\frac{1}{r^n}}$
terms are polynomials of suitable dimensions in $s^3,M,\ell^2$.
A closer inspection reveals that the limit $M\rightarrow 0$ is mathematically
very special.
The metric components $e^{\nu(r),\lambda(r)}$ acquire in this limit
their full analytical form, leading to the exact solution and very simple solution
\begin{equation}
	\fbox {$\displaystyle{ds^2=-\frac{r^2}{\ell^2}dt^2
	+\frac{\ell^2 dr^2}{\displaystyle{r^2+\frac{8s^3}{r}}}
	+r^2 d\varphi^2}$}
	\label{smetric}
\end{equation}
provided the associated scalar field $\phi(r)=vf(r)$ obeys the first-order
differential equation
\begin{equation}
	r(r^3+8s^3)f^\prime (r)-(r^3-4s^3)f(r)-r^3=0 ~.
	\label{f(r)}
\end{equation}
It can be easily verified that, with eq.(\ref{f(r)}) satisfied,
the two remaining second-order differential field equations get
automatically respected,
\textit{irrespective of the particular solution chosen}.

The special case $s<0$ catches our attention (note that the conceptually
different $s>0$ case, in particular the $s\rightarrow +0$ limit, has been
discussed in the literature in the context of maximal entropy packing
\cite{Spacking}).
The metric eq.(\ref{smetric}) is apparently singular at $r=2|s|$,
where even the scalar density
$\displaystyle{\sqrt{-g}=r^2\left(r^2+\frac{8s^3}{r} \right)^{-1/2}}$
happens to explode. 
But this turns out to be an artifact, as can be deduced from the
values of the various curvature scalars
\begin{eqnarray}
	&\displaystyle{{\cal R}=\frac{6}{\ell^2}~,~~{\cal R}^{\mu\nu}{\cal R}_{\mu\nu}
	=\frac{12(r^6+8s^6)}{\ell^4 r^6}~, }&\\
	&\displaystyle{{\cal R}^{\mu\nu\lambda\sigma}
	{\cal R}_{\mu\nu\lambda\sigma}
	=\frac{12(r^6+32s^6)}{\ell^4r^6} ~,}&
\end{eqnarray}
which stay all finite and continuous on the $r=2|s|$ surface.
Moreover, with eq.(\ref{f(r)}) satisfied, this surface stays non-singular
even in the Einstein frame where $g_E^{\mu\nu}=\phi^{-2}g^{\mu\nu}$.
For $s\neq 0$, the real curvature singularity is solely located at the origin.
While the overall situation partially reminds us of a black hole, it is quite
obvious that the $r=2|s|$ surface can be anything but an event horizon.
The $r=2|s|$ surface is characterized by the harmless single signature
change $(-,+,+)_{out} \leftrightarrow (-,-,+)_{in}$, the
nature of which we now attempt to reveal.

Focusing on the $s<0$ case, we find it convenient to set $s=-q^2$
from this point on.
Starting in the $r\geq2q^2$ region, eq.(\ref{f(r)}) admits there the exact
analytic solution
\begin{eqnarray}
	& \displaystyle{f_{+}(r)=\frac{(r^3-8q^6)^{1/2}}{r^{1/2}}
	\int_r^{\infty}\frac{x^{5/2}dx}{(x^3-8q^6)^{3/2}}=}& \nonumber\\
	& \displaystyle{\frac{5r^{3}_2 F_1\left(
	\textstyle{-\frac{1}{2},\frac{1}{3};\frac{4}{3};\frac{8q^6}{r^3}}\right)
	-2(r^3+4q^6)_2F_1\left(
	\textstyle{-\frac{1}{3},\frac{1}{2};\frac{4}{3};\frac{8q^6}{r^3}}\right)
	 }{3r^{3/2}(r^3-8q^6)^{1/2}}}&
	 \label{fplus}
\end{eqnarray}
The upper limit of the integration interval is nothing but the choice
$r_0\rightarrow\infty$ applied earlier in eq.(\ref{asymptotic}).
A technical difficulty is that the hypergeometric function $_{2}F_{1}(a,b;c;z)$
has a branch cut discontinuity in the complex z-plane running from 1
to $\infty$.
This means that the \textit{independent} solution $f_{-}(r)$ of eq.(\ref{f(r)}),
at the region $r\leq 2q^{2}$, must be stitched to $f_{+}(r)$ precisely
on the $r=2q^{2}$ surface.
Eq.(\ref{f(r)}) is a first order differential equation, and as such, its solutions
can be stitched at almost any point just on continuity arguments.
In our case, however, the junction value
\begin{equation}
	f_{+}(2q^{2})=f_{-}(2q^{2})=\frac{2}{3}
\end{equation}
is dictated by the differential equation eq.(\ref{f(r)}) itself, and is not
subject to any boundary conditions.
To see the point, notice the approximate expansions
\begin{eqnarray}
	f_{+}(2q^{2}+\epsilon) 
	\simeq \frac{2}{3}+k_{+}\sqrt{\epsilon} ~,~~
	f_{+}^{~\prime}(2q^{2}+\epsilon)
	\simeq \frac{k_{+}}{2\sqrt{\epsilon}}	~,
	\label{expandplus} \\
	f_{-}(2q^{2}-\epsilon) 
	\simeq \frac{2}{3}-k_{-}\sqrt{\epsilon} ~,~~
	f_{-}^{~\prime}(2q^{2}-\epsilon)
	\simeq \frac{k_{-}}{2\sqrt{\epsilon}} ~.
	\label{expandminus}
\end{eqnarray}
We note in passing that a perfectly smooth transition would require
$k_{\pm}=0$, with the price being $\phi(r)\sim r$ at large distances.
Insisting, however, on $\phi(r)\sim v$ at spatial infinity, the exact 
value $k_{+} =\sqrt{\frac{\pi}{3}}~ \frac{\Gamma\left(\frac{4}{3}\right)} 
{\Gamma\left(\frac{5}{6}\right)} \simeq 0.572$ can be calculated directly
from eq.(\ref{fplus}).
The fact that $\phi_{,r}$ diverges like $\epsilon^{-1/2}$ is not alarming,
given that $g^{\mu\nu}\phi_{,\mu}\phi_{,\nu}\simeq
\pm \frac{3v^{2}q^{2}}{2\ell^{2}}k_{\pm}^{2}$ is finite.
By the same token, $\phi_{;rr}$ diverges like $\epsilon^{-1}$, but as is
evident from the Klein Gordon equation, $g^{\mu\nu}\phi_{;\mu\nu}$
is finite and continuous.
While the value of $k_{-}$ has not been fixed at this stage, notice
that the metric eq.(\ref{smetric}) is independent of the particular choices
of $k_{\pm}$.
Moreover, as we shall soon see, it is eq.(\ref{smetric}) that tells us
that the two $\pm$-regions do not communicate geodesically. 

\begin{figure}[h]
	\includegraphics[scale=0.45]{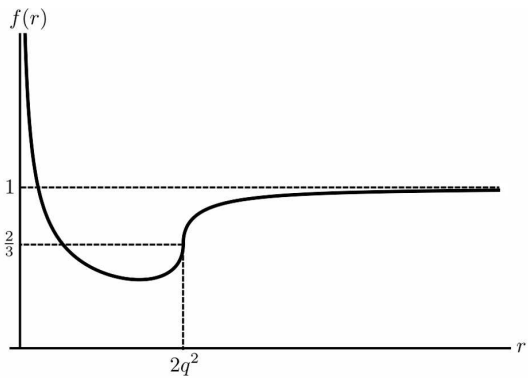}
	\caption{The (normalized) reciprocal Newton constant $f(r)$, is
	plotted as a function of the circumferential radius $r$.
	The asymptotic behavior $f(r)\sim 1$, and the matching at the
	gravitational mirror are emphasized.}
	\label{fig1}
\end{figure}

Choosing $k_{+}=k_{-}$, we can calculate $f_{-}(r)$, and find
\begin{equation}
	f_{-}(r)=\sqrt{\frac{8q^6}{r}-r^2}\left( \frac{c}{2q^2}
	+\int_0^r\frac{x^{5/2}dx}{(8q^6-x^3)^{3/2}}\right)
	\label{fminus}
\end{equation}
where $\displaystyle{c=\frac{\sqrt{\pi}}{9}
\left(\frac{\Gamma(\frac{1}{6})}{\Gamma(\frac{2}{3})}
-\frac{\Gamma(\frac{1}{3})}{\Gamma(\frac{5}{6})}\right)}\simeq 0.342$.
The combined $f(r)=f_{+}(r)+f_{-}(r)$, the reciprocal (normalized) effective
Newton constant, is depicted in fig.\ref{fig1}.

Back to the metric eq.(\ref{smetric}), we now reveal the special physical
features which characterize the $r=2q^2$ surface.
As hinted by the associated signature change $(-,+,+)\longleftrightarrow (-,-,+)$,
it is worthwhile studying first null geodesics.
The physical properties of light ray trajectories solely depend on the ratio
\begin{equation}
	1-\eta=\frac{L^2}{E^2\ell^2}\geq 0 
\end{equation}
of the conserved angular momentum $L=r^2 \dot{\varphi}$ to the conserved
energy $E=\ell^{-2}r^2 \dot{t}$.
Conveniently normalizing the affine parameter $\lambda$, the null geodesics
obey
\begin{equation}
	\left(\frac{dr}{d\lambda}\right)^2
	-\eta\left(1-\frac{8q^6}{r^3} \right)=0~.
\end{equation}
The equivalent mechanical problem involves the effective  potential
$\displaystyle{V_{eff}=\eta\left(\frac{8q^6}{r^3}-1\right)}$, and a vanishing
total mechanical energy.
Depending on the sign of $\eta$, the discussion trifurcates:

\smallskip
\noindent $\bullet$ $\eta>0$ null geodesics can only live at the outer region
	$r > 2q^2$.
	From the point of view of an external observer, the $r=2q^2$ surface
	acts like a white mirror, reflecting all incident null geodesics, thus
	resembling 't Hooft's brick wall, at least in the
	Susskind-Lindesay interpretation \cite{brick}.

\smallskip
\noindent $\bullet$ $\eta<0$ null geodesics, on the other hand, can only survive at the
	inner region $r <2q^2$.
	Null geodesics hitting the $r=2q^2$ surface from the inside get
	reflected, and are doomed to spiral into the singularity at the origin.
	The light cone structure excludes radial internal geodesics.

\smallskip
\noindent $\bullet$ This leaves $\eta=0$ to tag the only closed null geodesic
	available, which must live of course on the mirror itself.
	In some sense, the mirror can be viewed as the locus of
	"frustrated" light rays.

To be contrasted with a black hole event horizon, which serves
as a one-way membrane, no light ray can cross the two-sided
gravitational mirror.
In particular, from the point of view of an external observer, the
$0\leq r \leq 2q^2$ core does not reveal any piece of information, and is
in fact black and furthermore impenetrable.
This way, gravitational mirroring offers a novel sealing off mechanism
of curvature singularities, above and beyond the protection \cite{censor} 
offered by black holes.
To be a bit more specific, an external observer can only probe a bounded
Kretschmann curvature ${\cal R}^{\mu\nu\lambda\sigma}
{\cal R}_{\mu\nu\lambda\sigma}\leq 18\ell^{-4}$. 

\begin{figure}[h]
	\includegraphics[scale=0.45]{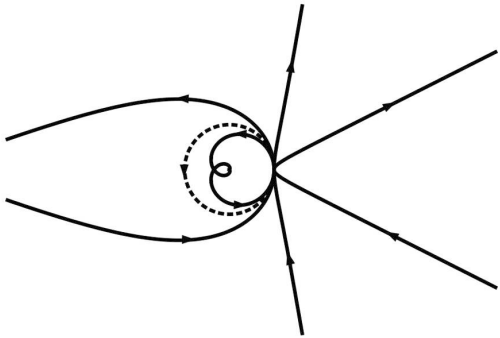}
	\caption{While external $(\eta>0)$ null geodesics get scattered,
	internal null geodesics $(\eta<0)$ are doomed to spiral into the
	singularity at the origin.
	The dashed $\eta=0$ concentric circular geodesic marks the
	two-sided gravitational mirror.}
	\label{fig2}
\end{figure}

The null trajectories $r(\varphi)$ are governed by the equation
\begin{equation}
	\frac{d\varphi}{dr}=\pm \frac{\ell}{r^2}
	\left[\frac{\eta}{1-\eta}\left(1-\frac{8q^6}{r^3}\right)\right]^{-1/2}~,
\end{equation}
a few of them are depicted in Fig.\ref{fig2}.
All null geodesics, without exception, are symmetric and cannot
escape touching the mirror, and unless $\eta=1$ (external radial),
they do it tangentially.

We now switch on the mass parameter $M$, and argue, bases on a
combined numerical and local analytic analysis, that the gravitational
mirror solution, while being correspondingly modified, is still there.
Starting from proper boundary conditions at large distances, that is
$\phi\rightarrow v$, see eq.(\ref{expansion}), the numerical graphs
exhibit the exact (verified numerically) mirror-like behavior just
outside some $r=h$, thereby implying, following the trail paved by
eqs.(\ref{expandplus},\ref{expandminus}), the tenable $\sqrt{\epsilon}$
expansion 
\begin{eqnarray}
	&& \nu(h+\epsilon) =
	\log\left(N_0+N_1 \sqrt{\epsilon}
	+N_2 \epsilon+...\right) ~, \nonumber \\
	&& \lambda (h+\epsilon) =
	-\log\left [\epsilon\left(L_0+ L_1 \sqrt{\epsilon}
	+L_2 \epsilon+...\right)\right ] ~,\\
	&& \phi(h+\epsilon)  = 
	\phi_0+\phi_1 \sqrt{\epsilon}
	+\phi_2 \epsilon+... ~. \nonumber 
	\label{Mexpansion}
\end{eqnarray}
Indeed, one coefficient is in charge of the general relativistic large $r$
behavior of the scalar field, two coefficients are used to extract $M$
and $s^3$ from the asymptotic expansion (note that
$N_{1},L_{1}\rightarrow 0$ as $M\rightarrow 0$), another coefficient
($N_0$) gauges the scale of time, and the rest of the coefficients can
be calculated order by order.
From the zeroth order, for example, we can calculate the location of the
mirror in terms of the above coefficients, namely
\begin{equation}
	h=\frac{1}{2}\ell^2 L_0 \left(3-\frac{v}{\phi_0}\right)^{-1}~,
\end{equation}
and recall the previous result $\phi_0\rightarrow \frac{2}{3}v$
as $M\rightarrow 0$, which assures $h>0$ at least for small enough
masses.
The crucial observation now is that all curvature scalars are in fact finite
at $r=h$. 
One representative example (the other are very lengthy) is given by
\begin{equation}
	{\cal R}(h+\epsilon)=\frac{L_1 N_1}{8 N_0}
-\frac{L_0 N_1^2}{8 N_0^2}+\frac{L_0 N_2}{2 N_0}+\frac{L_0}{h}
+{\cal O}(\sqrt{\epsilon})~.
\end{equation}
This allows for matching the just outside with the just inside
expansions, in the spirit of eqs.(\ref{expandplus},\ref{expandminus}),
and in accord with the matching procedure \cite{SignatureChange},
thereby constituting a non-singular \emph{massive} two-sided mirror.
The analysis can be easily extended to an asymptotically flat 4-dim
spacetime, with the main difference being the Yukawa suppression
of the scalar field effect at large distances  \cite{4dim}. 

Finally, digesting the latter conclusion, we return to the original
massive spinless BTZ black hole configuration eq.(\ref{BTZ}), and switch
on a tiny negative scalar charge, with $q^2 \ll \sqrt{M\ell^2}$.
No matter how small $q^{2}>0$ is, a phase transition takes place.
The non-singular horizon surface of radius $\sqrt{M\ell^2}$ becomes
geodesically sealed from both sides, and transforms into a non-singular
two-sided gravitational mirror of a slightly larger radius
\begin{equation}
	h\simeq \sqrt{M\ell^2}
	\left( 1+\frac{4q^6}{(M\ell^2)^{3/2} }\right)~.
\end{equation}
The bottom line is that one black object has replaced the other. 
Thus, the fact that a black hole is Nature's ultimate information storage
makes us wonder whether any information leakage has occurred during the
phase transition.
If the answer to this question is (even partially) negative, then the two-sided
gravitational mirror must carry some entropy.
On holographic/geometric grounds \cite{holographic}, one may even expect
the amount of this entropy to be at most half the circumference of the mirror
$S\leq \pi h$.
At this stage, however, we cannot support nor falsify such a possibility
by means of a field theoretical calculation.

\acknowledgments{We would like to thank BGU president Prof. R. Carmi
for the kind support.}

\end{document}